\documentclass[aps,prc,showpacs,showkeys,floatfix,nofootinbib,twocolumn,
10pt]{revtex4-1}
\usepackage[dvips]{graphics,graphicx}
\usepackage[colorlinks=true,linktocpage=true,linkcolor=blue,citecolor=blue]{hyperref}
\usepackage[usenames,dvipsnames]{color}
\usepackage{amsmath, amssymb}
\usepackage{multirow}
\usepackage{longtable}
\usepackage{color}
\usepackage{hyperref}
\usepackage{cleveref}
\usepackage[normalem]{ulem}  
\usepackage{float}
\usepackage{graphicx}
\usepackage{epsfig}
\usepackage{bm}
\usepackage{amsthm}
\usepackage[caption = false]{subfig}

\begin{document}

\preprint{}

\title{ Analyzing COVID-19 pandemic with a new growth model for population ecology }

\author{Deeptak Biswas}
\affiliation{Department of Physics, Center for Astroparticle Physics \& Space Science
Bose Institute, EN-80, Sector-5, Bidhan Nagar, Kolkata-700091, India }
\email{deeptak@jcbose.ac.in} 

\author{Sulagna Roy}
\affiliation{Department of Botany, Santipur College, N.S Road, Santipur, Nadia, West Bengal 741404, India}
\email{beingsulagna@gmail.com}

\date{\today}

%
\begin{abstract}
We have proposed a new form of growth rate for population ecology. Generally, the growth rate is dependent on the size of the population at that particular epoch. We have introduced an alternative time-dependent form of growth rate. This form satisfies essential conditions to represent population growth and can be an alternative form for growth models to analyze population dynamics. We have employed the generalized Richards model as a guideline to compare our results. Further, we have applied our model in the case of epidemics. To check the efficacy of our model, we have verified the 2003 SARS data. This model has estimated the final epidemic size with good accuracy. Thereafter, we intend to describe the present COVID-2019 pandemic. We have performed our analysis with data for Italy, Spain, and Germany. Following, we have tried to predict the number of COVID-19 cases and the turning point for the USA, UK, and India. 
\end{abstract}

\keywords{
Population ecology, Growth model, Richards model, SARS, COVID-19
}
\pacs{87.23.Cc,  92.20.jm}
\maketitle

\section{Introduction}
The time evolution of population growth is fundamentally interesting from both the point of ecology and physics \cite{Turchin}. Certain factors and limiting conditions determine the trajectory of a population density. The growth rate seems to depend on the population size and follows some general differential equations. Various epidemic diseases also follow the same dynamical pattern \cite{WANG201212, hsieh2004sars, hsieh2006real, hsieh2009intervention, CHOWELL201671}. One can predict the progress of an epidemic and the cumulative number of cases by studying these models, which may provide insights into ongoing outbreaks to expedite public health responses.
 
Recently a pandemic condition has emerged, causing severe pneumonia
and respiratory disorder. The first cluster of viral pneumonia cases of unknown cause was reported on 31st December 2019 at Wuhan, Hubei, China. On 30th January 2020, WHO first declared COVID-19 as a serious public health concern, and then on March 11th it was recognized as a pandemic \cite{WHOPAN}. The International Committee on Taxonomy of Viruses (ICTV) identified this virus as "Severe Acute Respiratory Syndrome Coronavirus 2" (SARS-CoV-2)\cite{of2020species}. This is giving rise to an enormous number of cases of pneumonia. Governments are attempting to exercise all measures and rapid movements to control this pandemic. In February 2020, cumulative cases in SARS–CoV-2 exceeded the previous SARS outbreak. These emerging infections have become a serious concern over the last two decades, with absolutely no specific control measures and proper medication. Recent studies have conferred COVID-19 to have similarities in disease patterns with previous SARS cases, \cite{huang2020clinical, HOFFMANN2020271, song2019sars}.

Mathematical models \cite{anderson1991infectious, onwubu2019application, tang2020updated, villela2020discrete, pell2018using, birch1999new, Batista2, WANG201212, hsieh2004sars, hsieh2006real, hsieh2009intervention} may be utilized to predict infectious disease spread. According to Keyfitz Ref.\cite{keyfitz1972future}, in the case of prediction, there is a difference between "projection" and forecast. Projection depends on some set of assumptions and the data sets of that particular epoch,  whereas the forecast is an unconditional statement of future trajectory. In a stochastic system like population growth and epidemics, these two seem to supervise each other. 

In epidemic studies, one can employ the growth equations to provide estimates of the cumulative number of cases, peak timing, infection rate, etc. These approaches allow us to predict the future dynamics of the epidemics, associated with the best fit to data.  A general logistic equation (first proposed in Ref. \cite{verhulst1838notice}) with a sigmoid shaped solution for cumulative number may have a good approximation of the final epidemic size and the turning point. Richards model Ref.\cite{Richards} further generalized this with an extra added parameter. The performance of the various logistic model and the generalized Richards model (GRM) (\cite{WANG201212} and references therein) can be calibrated with available epidemic data. GRM's phenomenological nature was capable to estimate the most accurate growth rate and final epidemic size \cite{pell2018using}. In a general case, there can be multiple waves of an epidemic, so implementing a multi-phase Richards model with multiple turning points depending on the interval may be important Ref.\cite{hsieh2009richards}.  
 
Following the eruption of COVID-19, several studies have made short-time predictions and forecasted the dynamics. Ref.\cite{Batista2} predicted the final cumulative numbers of cases from both the logistic and SIR (Susceptibles  Infectives Removed) model. Ref.\cite{roosa2020real} has investigated cumulative numbers of cases from the early data of Hubei, China. In the case of an imposed lockdown, there will be local maxima and minima in daily outbreaks, where an SEIR-type (Susceptible-Exposed-Infectious-Recovered) type model may fail. Ref.\cite{fodor2020differential} has discussed this. Recently Ref.\cite{sabyada1, sabyada2} has tried to investigate the importance of lockdown and difference in sigmoid and exponential like growth for various states of India. 

The common perception from all the studies is that growth rate and various estimated parameters are highly dependent on the available data. So with possible uncertainties, the predictions from these models may or may not match at the end of the pandemic. But applying these models may capture the essence of the outbreak via short time predictions and help to take preventive steps. These are also helpful in government interventions and to impose lockdown, which may reduce the number of cases and fatalities.
 
In this manuscript, we have introduced a different approach to describe the growth rate. Rather than relating the growth rate to population size, we have advocated a simple time-dependent form of growth rate equation. Our method seems to have all the necessary elements to describe a general population growth and has delineated the previous SARS-2003 epidemic data suitably. Further, we have analyzed data of COVID-19 for Italy, Germany, and Spain as a benchmark study and then tried to predict the future scenario in the case of the USA, the UK, and India. We have used the Generalized Richards Model (GRM) as a guideline for these results. There are beautiful agreements between GRM and our model estimation. This simple form of growth rate can be a viable alternative for various ecological growth models.
   
We have reviewed the logistic model and GRM in section \ref{sc.growth} and described our proposed form of equations in section \ref{sc.our}. In section \ref{Results} we have explained our results and finally, we have summarized in \ref{summary}. 

\section{Growth models in population ecology}\label{sc.growth}
The growth of a population can be represented by some general laws \cite{PELL201862}. In the canonical picture, one can find a pattern for the growth of a population. That is, 
\begin{equation}
\frac{dN}{dt}\approx rN
\end{equation} 
Here, $r$ determines the relative growth rate or growth rate per capita. $N$ is the population size, and $dN/dt$ is the rate of change. The growth of a population depends on general constraints like availability of resources, environmental conditions, birth rate, death rate, etc. The growth rate ($r$) becomes negative when the population size $N$ crosses a threshold value ($K$).

\subsection*{Generalized Richards Model (GRM)}
In the standard logistic model proposed by Verhulst (1838), the growth rate of a population is proportional to the size of the population and constrained by a maximum number of population ($K$). Let $N(t)$ represent the population size at time $t$. In the logistic model, this evolves as follows (Ref.\cite{WANG201212} and references therein):
\begin{equation}\label{log}
\frac{dN(t)}{dt}=r N(t)\left[1-\frac{N(t)}{K}\right]
\end{equation}
Here, $r$ is the growth rate and $K$ is the carrying capacity. Richards generalized this model with one additional parameter $a$ \cite{Richards}.
\begin{equation}\label{GRM}
\frac{dN}{dt}=\tilde{r} N(t)\left[1-\left(\frac{N(t)}{K}\right)^{a}\right]
\end{equation}
$a$ measure the deviation from the standard logistic model. In Ref.\cite{WANG201212}, the exponential term has been shown to have a relation with the basic reproduction number $R_0$, which is defined as the expected number of cases directly generated by one case in a population.
The solution of Eq.\ref{GRM} is,
\begin{equation}\label{solgrm}
N(t)=K\left[1+a e^{- a \tilde{r} \left(t-t_{p}\right)}\right]^{-1 / a}
\end{equation}
where $t_{p}$ is the turning point defined as the time when the rate $dN(t)/dt$ becomes maximum. 
We can evaluate the time ($t_p$) from, 
$$
\frac{d}{dt}(dN/dt)=0
$$ 
The Generalized Richards Model (GRM) was initially introduced in the context of ecological population growth. Recently this model has been studied in epidemiology for real-time prediction of diseases \cite{hsieh2004sars, hsieh2006real, hsieh2009intervention}. We shall use this model as a guideline in our study for the ongoing pandemic COVID-19. For simplification, we have done a rescaling $r=a \tilde{r}$, following Ref.\cite{WANG201212}. We shall use the rescaled form of cumulative cases Eq.[\ref{solgrm}].
 
\section{New approach}\label{sc.our}
In the case of epidemics, one can relate $dN/dt$ with the number of daily infected persons, i.e new informed cases per day. From data of daily infections \cite{worldometer}, it seems that the general trend of $dN/dt$  for COVID-19 follows a skewed pattern. The general perception is that rapid growth is followed by an exponential decay. In standard population ecology, a general rate equation depends on the cumulative population $N(t)$ of that particular time. Here we have taken a different approach and investigated whether one can study this growth pattern only as a function of time.

We have performed normalization of daily new case data with the maximum value, which is $(dN/dt)/(dN/dt)_{max}$. The general trend of this normalized increment has similarity with chi-distributions, where the general form is,
\begin{equation} \label{chi}
f(x ; k)=\frac{x^{k-1} e^{-x^{2} / 2}}{2^{k / 2-1} \Gamma\left(\frac{k}{2}\right)}
\end{equation}
here $\Gamma(z)$ is the gamma function. This distribution has one parameter $k$, which specifies the degrees of freedom. The denominator of Eq.[\ref{chi}] corresponds to normalizing factor.

Noticing this trend, we intend to propose a new growth rate equation. A Maxwell–Boltzmann distribution gives rise to a symmetric distribution of $dN/dt$ with time. But we have noticed that in most of the cases, the tail of the curve is stretched. A right-skewed distribution pattern has been observed for most of the countries, which does not follow a gaussian structure. So rather than assuming a particular structure, here we propose a generalized rate of infection $dN/dt$,
\begin{equation}\label{DS}
\frac{dN}{dt}=f(t)=t^{m} e^{-(t/\lambda)^{n}}
\end{equation}
$f$ is a funtion of time. To understand the important of these paramters, let's look at the derivative of Eq.\ref{DS},
\begin{equation} \label{DSeq}
\frac{dN^\prime }{dt}=\frac{N^\prime}{t} m [ 1-\frac{n}{m}(t/\lambda)^n]
\end{equation}    
We can find the time when the infection rate $\frac{d N}{d t}$ reaches its maximum point from Eq.[\ref{DSeq}]. At this point $\frac{d^2 N}{dt} = 0$. In our model this time ($t_p$) is,
\begin{equation}\label{tp}
t_p=\lambda (\frac{m}{n})^{1/n}
\end{equation} 
So $t_p$ is proportional to $\lambda$. With this form of $t_p$, Eq.\ref{DSeq} becomes,
\begin{equation} \label{DSeqn}
\frac{dN^\prime }{dt}=\frac{N^\prime}{t} m [ 1-(t/t_p)^n]
\end{equation}    
If $t$ is much smaller than $t_p$, the rate will be dominated by the factor $m$ and the rate will be proportional to $t^m$ i.e $dN/dt \approx t^m$. Whereas, for $t \gg t_p $ the fall of the growth rate is controlled by the parameter $n$. So at later part of the pandemic $dN/dt \approx exp(-t^n)$. In a pandemic disease like COVID-19, one should expect a larger value of $m$ than $n$, as the infection spreads rapidly and sustains for a long period.

One can estimate the cumulative number cases $N(t)$ in any time by integrating out $dN/dt$ for any $ t$. In our cases, an exact analytical form could not be obtained, except for the asymptotic limit $t \rightarrow \infty$. We have performed numerical integrations to find the number at any given $t$. The growth rate will be $0$ at $t \rightarrow \infty$, so no new cases will be informed and the cumulative number of cases for the pandemic can be found as,           
\begin{equation} \label{Nassy}
N_{total}=\int_{0}^{\infty} t^{m} e^{-(t/\lambda)^{n}} d t=\frac{1}{n} \lambda^{m+1} \Gamma\left(\frac{m+1}{n}\right)
\end{equation}
This $N_{total}$  correspondence to carrying capacity ($K$) of Generalized Richards Model of population growth and can be assumed to be the size of the epidemic. With the above formalism, we shall try to analyze the data of SARS-2003 and COVID-19. Efficacy of our model can be checked along with the generalized Richardson model, via prediction of both $N_{total}$ and $t_p$.

\section{Result and Discussion}\label{Results}
\subsection*{Data analysis and formalism}
First, we shall employ our model and GRM in data for the SARS-2003 epidemic case.  Studies on virology and epidemiology of COVID-19 establish it as a zoonotic \cite{millan2020new}, especially bat as an initial host. The previous epidemic SARS 2003 also belongs to the same natural reservoir \cite{song2019sars}. Clinical studies on patients also have established a close resemblance of COVID-19 with SARS \cite{huang2020clinical}. SARS CoV was transmitted globally and was declared an epidemic. Therefore, investigation with SARS data \cite{WHOSARS, Kaggle} will act as a benchmark study for the present pandemic. We shall predict the turning point $t_p$ ( maximum growth rate ) and the cumulative number of cases in SARS-CoV, from both models. A satisfying agreement with the prediction from GRM will check the efficiency of our proposed rate equation. Then we shall apply our model in case of the ongoing pandemic COVID-19.  

In case of GRM model, the $dN/dt$ is not an explicit function of $t$, whereas the solution has a form of $f(t)$. So the $t-dependent$ behavior can be investigated from the cumulative data set. We have fitted cumulative data of SARS with the GRM model and found the parameters $K, a, r$, and $t_p$. Thereafter we employ these parameters to get the functional form of $dN/dt$. Whereas, with our model, we have parameterized the data of daily new cases and predicted $n, m$, and $\lambda$. As there is no analytical solution of $N(t)$ from our model, we have to perform numerical integrations to evaluate cumulative $N(t)$. An immediate crosscheck about the accuracy of our integration result can be done by performing $\int Ndt$ in the limit $[0, \infty]$ and employing Eq.[\ref{Nassy}]. Further, we have compared results for cumulative cases in both the model and verified them with corresponding data.

\subsection*{Verification with SARS data}
\begin{figure}[H]
\includegraphics[width=.50 \textwidth]{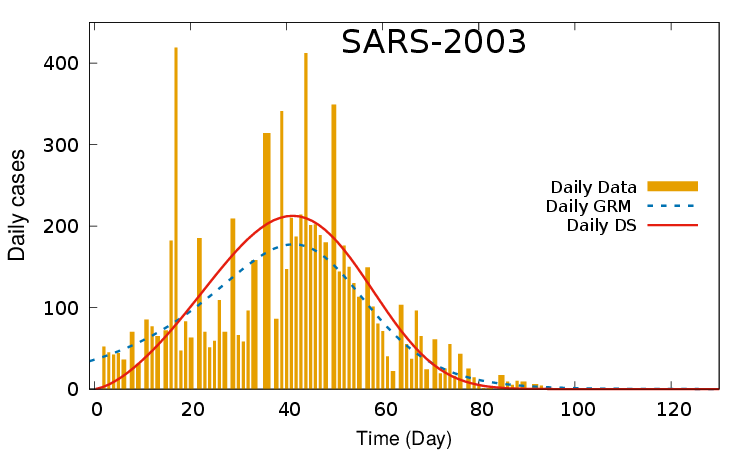}
\caption{Day vs Increment rate in SARS-2003. Histogram denotes daily informed new cases. Blue dotted lines are GRM fitted curve. The red continuous line is fitted with the present study.}
\label{fg.sars1}
\end{figure}

We have analyzed SARS data for $90$ days from 17$^{th}$ March to 17$^{th}$ June of 2003. We have parameterized our model with data of daily informed new cases, whereas the RGM has been fitted using the cumulative data. In Fig.[\ref{fg.sars1}], we have plotted daily increment of SARS-2003 cases. Though our fitted curve (red line) has little difference with that of RGM (blue dashed), both seem to replicate the data (orange bars) quite suitably. 

\begin{figure}[H]
\includegraphics[width=.50 \textwidth]{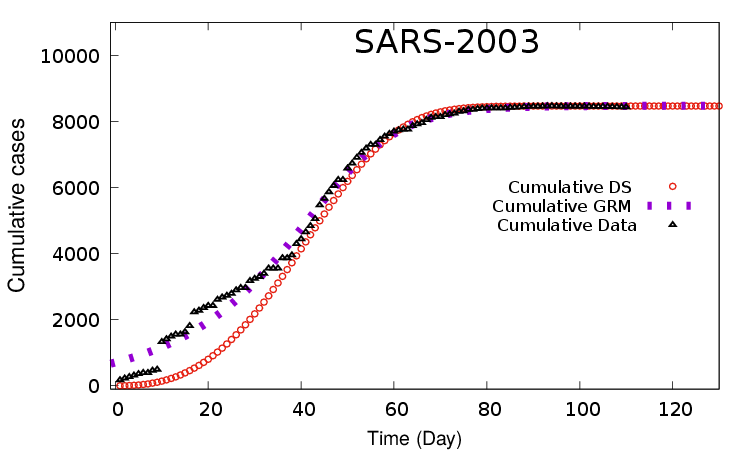}
\caption{Day vs Cumulative number cases for SARS-2003. Black triangles denote data. magenta dashed lines are GRM fitted curves. The red continuous line is predicted with the proposed model.}
\label{fg.sars2}
\end{figure}

Next, we try to predict cumulative data from our model by performing numerical integration. As the saturation trend originates around $80$ days, we have evaluated the cumulative number for a $120$ days time period. Though there is disagreement in the initial phase of the cumulative curve, our model has suitably reproduced the data in the later phase of the epidemic (once it reaches $t_p$) in Fig.[\ref{fg.sars2}]. This disagreement may result from the peak of day 17 of the increment curve.    
There is a satisfying agreement in case of the predicted value of $t_p$ and $N_{total}$, between our model and RGM. The estimated values are listed in table [\ref{sarstable}]. Here we want to emphasize that though we do not have an exact analytical solution for cumulative number, we have satisfactorily reproduced the data. So our model can be a suitable alternative to discuss epidemic size and duration time.

\begin{center}
\begin{table}
\caption{Evaluated values of $t_p$ and $N_{total}$ for SARS-2003.}

\begin{tabular}{|c|c|c|c|c|}
\hline Model & $t_p$  & $N_{total}$   \\
\hline GRM   & 41.468 & 8475.208   \\
\hline This study   & 41.272  & 8467.912   \\
\hline Data  &		 NA       & 8434   \\

\hline
\end{tabular}
\label{sarstable}
\end{table}
\end{center}
\subsection*{Application in COVID-19 }
With the above formalism, we shall discuss the ongoing pandemic of COVID-19. We have utilized the data for countries that have crossed 1 Lakh in total cases. We have used data for Italy, Spain, and Germany in our initial study, with reported data as of 20$^{th}$ April 2020. Then we shall try to predict the scenario for the USA, UK, and India. 

There is diversity in the initial growth pattern among all the countries. This difference may depend on their geographical scenario, environment, and infection rate \cite{sajadi2020temperature}. But we have noticed that the general behavior is identical for all the cases, once the number of daily reported cases ($dN/dt$) reaches around $100$. We have counted that day as Day $1$ in our analysis. Data are available from February 15$^{th}$ in \cite{worldometer} for all countries. But depending on the above-mentioned condition we have used data from the last week of February for Italy and from the first week of March for most of the countries.      

\begin{figure}[H]
\subfloat[]{
{\includegraphics[scale=0.7]{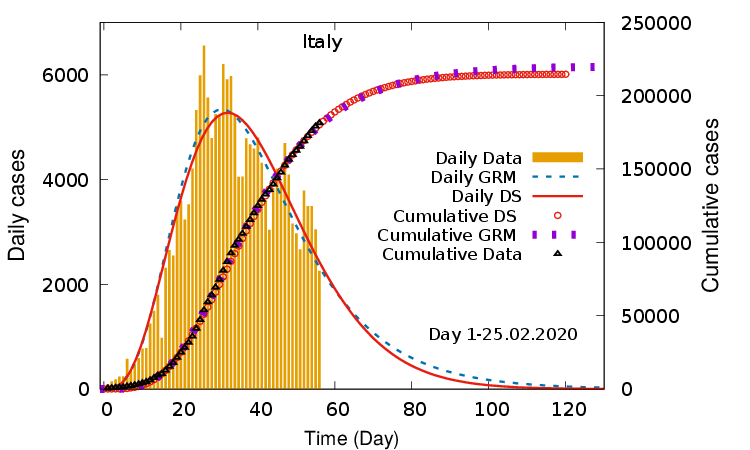}}
\label{fg.italy}
}\\
\subfloat[]{
{\includegraphics[scale=0.7]{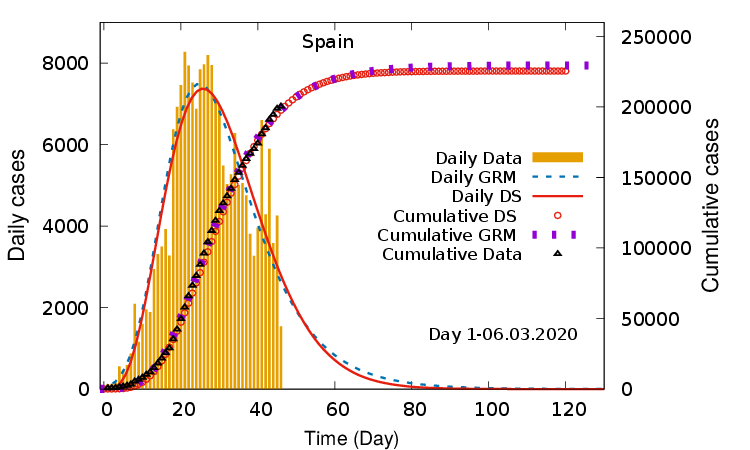}}
\label{fg.spain}
}\\
\subfloat[]{
{\includegraphics[scale=0.7]{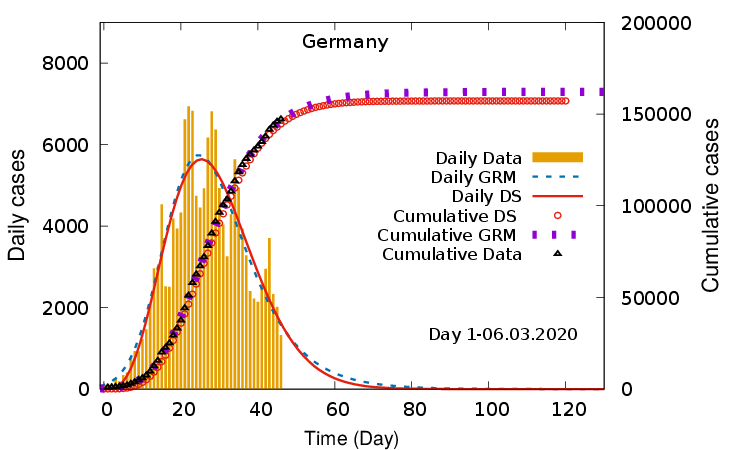}}
\label{fg.germany}
}
\caption{
\label{fg.eu}
(colors online) Plots for day vs. daily cases and cumulative cases in COVID-19 for Italy, Spain and Germany. Legends are same as mentioned in Fig.[\ref{fg.sars1}~~\ref{fg.sars2}]. Data are from \cite{worldometer}. Y axis in the left side denotes number of daily reported cases ($dN/dt$) and right Y axis denotes cumulative cases. For clarity we have plotted both cumulative and daily cases in same graph. X axis is the number of days. Day $1$ denotes the day when daily reported case ($dN/dt$) was around $100$ for first time. 
}
\end{figure}

In Fig.[\ref{fg.eu}] we have shown results for Italy, Spain, and Germany. These countries are important to understand the behavior of the COVID-19 pandemic in Europe. For all these cases, our model estimation (red continuous line) seems to reproduce data (orange bar) suitably and matches with that of the GRM (blue dashed line). The growth rate increases as day lapses and reaches its maximum value around day $35$ of our analysis. We have listed the value of turning point $t_p$ from both the models in table [\ref{tb.tpeu1}]. Turning point determines the day when the growth rate ( $dN/dt$ ) reaches its maximum value. Onwards this point the growth rate decreases and in the asymptotic limit, it becomes zero. 
\begin{center}
\begin{table}
\caption{Evaluated values of $t_p$ COVID-19.}

\begin{tabular}{|c|c|c|c|c|c|c|}
\hline Model      & Italy & Spain  & Germany  \\
\hline GRM        & 30.59 & 24.76  & 24.71   \\
\hline This study & 32.11 & 25.75  & 25.33   \\
\hline
\end{tabular}
\label{tb.tpeu1}
\end{table}
\end{center}
In our model, the parameter $m$ determines the initial phase and controls the growth. A larger value of $m$ is responsible for a steeper rise. On the other hand, $n$ determines the rate of decrease in the growth rate in the later phase. In all the cases, there is a similarity among the parameter sets (listed in table \ref{tb.DS}) and $m$ is almost three times higher than $n$. One can also infer that from the trend of the daily reported case. There is a faster increase in the number of cases, whereas the tail is extended. But in the case of GRM, the fitted value of $r$ and $a$ (listed in table \ref{tb.GRM}) has variation among these three countries. So the relative growth rate and basic reproduction number are different. Whereas the $m$ and $n$ have similar values in our model. This similarity may have interesting consequences. This may indicate that the growth pattern has an intrinsic time dependence, depending on environmental conditions and demography. Further studies with data from different regions of the globe may enlighten in this direction. 

We have also estimated the epidemic size ($N_{total}$) from our model. This should match with the carrying capacity $K$ of GRM, as both determines the final cumulative number of cases. We have seen a good agreement in Fig.[\ref{fg.eu}] between the predictions of the total number of cases. Our model successfully reproduced the total number of cases for SARS-2003. Here we have estimated the $N_{tot}$ as a prediction from our model. In all these cases, the pandemic seems to decelerate after $80$ days ($80$ days should be counted from day $1$ of our method i.e first day to report 100 new cases). Our model predictions have good agreement with the available cumulative number, so we can further utilize this model to predict growth rate and final epidemic size in the future. 

We have predicted cumulative numbers and $t_p$ for the USA, the UK, and India from the available data set as of 23$^{rd}$ April. Both models (GRM and this study) have good agreement with the present data and have made similar predictions. The daily reported case reached $100$ for the first time on 6$^{th}$, 12$^{th}$, and 23$^{th}$ March respectively. So in the case of India, the pandemic is still at the growth phase. In the case of the USA and the UK, it seems that the growth curve has reached its peak. We have listed our evaluated $t_p$ in table \ref{pretp}. In the USA and UK, the pandemic is expected to reach the asymptotic limit in 90 days from our day-1, with cumulative cases 13lakh and 2lakh respectively.

For India, GRM has predicted the date for the peak ($t_p$) is around 5$^{th}$ May, whereas our model has predicted around 8$^{th}$ May. This slight difference has an impact on the final cumulative number of cases. GRM has estimated around 140000 total cases on 250$^{th}$ day, whereas our model has determined the number to be around 130000. This difference of 10000 in the predicted total number of cases may be related to the fitting procedure. In the case of GRM, we have performed the fitting with the cumulative data, whereas for our model we have utilized the daily data of newly reported cases.

\begin{center}\label{pretp}
\begin{table}
\caption{Predicted values of $t_p$ COVID-19 for USA, UK and India. $t_p$ sould be counted from day 1, listed in last row.}

\begin{tabular}{|c|c|c|c|c|c|c|}
\hline Model      & USA & UK & India  \\
\hline GRM        & 34.67  & 31.97  & 47.67   \\
\hline This study & 37.19  & 32.84  & 44.13   \\
\hline Day 1      & 06.03.2020 & 12.03.2020  & 23.03.2020 \\
\hline
\end{tabular}
\label{covtp_table1}
\end{table}
\end{center}
     
\begin{table}
\caption{Estimated parameters from our model.}

\begin{tabular}{|c|c|c|c|c|c|c|}
\hline Country      & $m$ & $n$ & $\lambda$  \\
\hline Italy        &       3.23370225 &  1.22124715 &  14.46834768 \\
\hline Spain	    &      3.65860059  &  1.22767396 & 10.58388817  \\
\hline Germany      &       3.35989366 &  1.51246424 & 14.9453637   \\
\hline USA          &       3.48673953 &  1.57025617 & 22.37677876  \\
\hline UK           &       3.01084791 &  1.56642347 & 21.64073827  \\
\hline India        &       3.15434579 & 0.69054516  & 4.89067862   \\
\hline
\end{tabular}
\label{tb.DS}
\end{table}

We especially want to mention that, the study with data of China could be really interesting in this context, as this was the epicenter of the outbreak. But there are some confusions regarding the data of China. On February 12$^{th}$, Hubei province reported nearly 15000 new cases in a single day. The growth rate tended to decline before this day and the maximum reported case was nearly 4000 on Feb 4$^{th}$. Our model could not fit this sudden rise with this limited number of parameters. Except for this particular case, our data seems to have a satisfactory agreement. In the future, we shall try to describe this kind of sudden rise, incorporating both cumulative data and increment rate in our model.    

\section{Summary and Outlook}\label{summary}
In this article, we have advocated a new form of the growth rate of population ecology. The proposed rate equation solely depends on time, with a combination of both rapid rise and exponential decay. This simple method may be useful to analyze the growth rate pattern to time, independent of the population size. Our model has estimated the turning point and carrying capacity with good precision. We have evaluated the number of cumulative cases by implementing numerical integration.  The elegance of our proposed rate equation remains in its simplistic form with only three parameters. This model can be a viable alternative to the growth model in general population ecology. 

Following the application of the ecological growth model in the estimation of epidemic cases, we have tried to utilize our data in the previous SARS-C0V case of 2003. Further, we have discussed the present pandemic COVID-19 with our model. We have analyzed data for Italy, Spain, Germany to check the efficacy of our model. We have performed parametrization with the generalized Richards model as a benchmark study for our model. Our proposed model has a satisfying agreement with the GRM model in evaluating the turning point and final epidemic size for all the cases. We have also predicted the USA, the UK, and India using both GRM and our method. The pandemic seems to stay in India for a longer period than other countries due to a slower growth rate initially and late timing of the turning point. Certain external conditions like lockdown, rapid testing, etc. may have effects on these growth patterns, but that is beyond the scope of our present study. 

In the case of GRM, the growth rate depends on the population size, which may be misleading in the initial phase of epidemic situations, as the number of the total reported case remains very low. Whereas in our model the growth rate is a simple function of $t$. So rapid predictions can be made from the number of newly reported cases. Shortly, we shall intend to encounter these issues with a further modified version of our model. This simple and elegant form of growth rate equation may be widely used for studying tumor cell growth, algal growth, etc.

\section*{Acknowledgements}

This work is funded by UGC and DST of the Government of India. We want to thank Dr. Jyotirmoy Pandit and Thirthangkar Biswas for their suggestions regarding plotting. We especially thank Samapan Bhadury, Sabyasachi Ghosh, and Sougata Guha for the critical reading of the manuscript. DB acknowledges Dr. Sudipto Muhuri for endorsing in arxiv biophysics. 

\bibliography{corona}

\begin{figure}[H]
\subfloat[]{
{\includegraphics[scale=0.7]{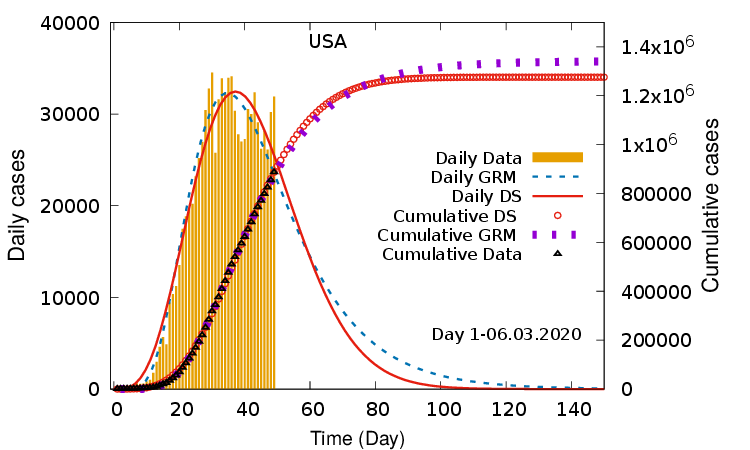}}
\label{fg.usa}
}\\
\subfloat[]{
{\includegraphics[scale=0.7]{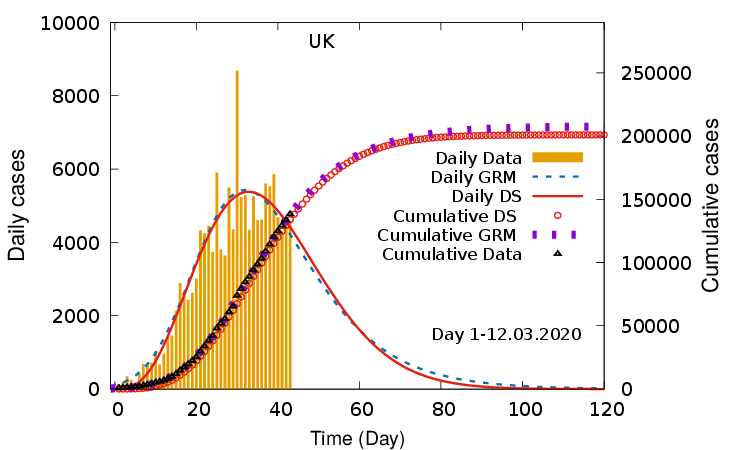}}
\label{fg.uk}
}\\
\subfloat[]{
{\includegraphics[scale=0.7]{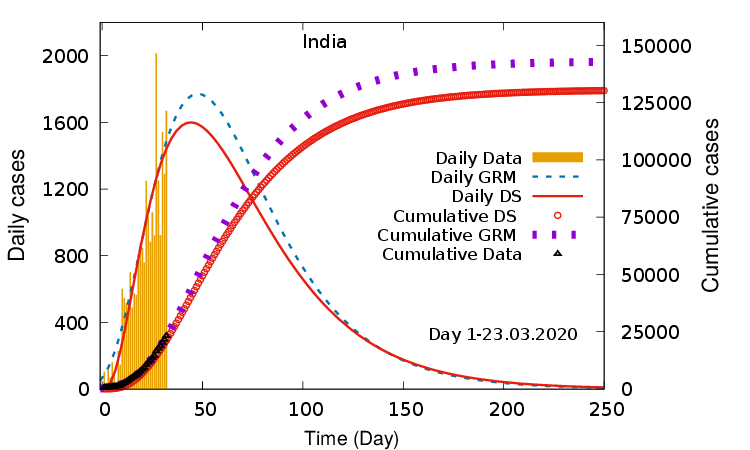}}
\label{fg.India}
}
\caption{
\label{fg.prediction}
(colors online) Predictions for day vs. daily cases and cumulative cases in COVID-19 for USA, Uk, and India. Legends are the same as mentioned in Fig.[\ref{fg.sars1}~~\ref{fg.sars2}]. Data are from \cite{worldometer}. Y-axis in the left side denotes the number of daily reported cases ($dN/dt$) and the right Y-axis denotes cumulative cases. For clarity, we have plotted both cumulative and daily cases in the same graph. The X-axis is the number of days. Day $1$ denotes the day when daily reported case ($dN/dt$) was around $100$ for the first time. 
}
\end{figure}

\onecolumngrid
\appendix
\begin{table}
\caption{Estimated parameters from GRM model.}

\begin{tabular}{|c|c|c|c|c|c|c|}
\hline Country      & $K$ & $a$ & $r$ & $t_c$ \\
\hline Italy        &     $2.20118104 \times 10^{+05}$ & $-9.38737066 \times 10^{-02}$ &  $6.28299920 \times 10^{-02}$ &  $3.05998285 \times 10^{+01}$ \\
\hline Spain	    &     $2.29514756 \times 10^{+05}$ &  $2.54206716 \times 10^{-02}$ &  $8.98775288 \times 10^{-02}$ &  $2.47666092 \times 10^{+01}$ \\
\hline Germany      &     $1.62109343 \times 10^{+05}$ &  $2.15841168 \times 10^{-01}$ &  $1.06587950 \times 10^{-01}$ &  $2.47109745 \times 10^{+01}$ \\
\hline USA         &      $1.34105575 \times 10^{+06}$ &  $-1.27100099 \times 10^{-01}$ &  $6.12501036 \times 10^{-02}$ &  $3.46779860 \times 10^{+01}$  \\
\hline UK          &      $2.07848200 \times 10^{+05}$ &  $1.79034139 \times 10^{-01}$ &  $7.72675411 \times 10^{-02}$ &  $3.19759960 \times 10^{+01}$ \\
\hline India       &      $1.42982935 \times 10^{+05}$ & $-1.20496716 \times 10^{-01}$ &  $3.16275515 \times 10^{-02}$ &  $4.76732500 \times 10^{+01}$ \\
\hline
\end{tabular}
\label{tb.GRM}
\end{table}

\end{document}